\documentclass{ifacconf}
\usepackage{graphicx}
\usepackage{natbib}
\usepackage{amsmath, amsfonts, amssymb}
\usepackage{algorithm} 
\usepackage{algpseudocode}
\usepackage{url}
\usepackage[textwidth=1.5in]{todonotes}
\usepackage{mathtools}

\newcommand*{\QEDA}{\hfill\ensuremath{\blacksquare}}
\DeclareMathAlphabet{\mathcal}{OMS}{cmsy}{m}{n}
\DeclareMathOperator{\sign}{sign}
\algnewcommand\And{\textbf{and}}

\newtheorem{defin}{Definition}
\newtheorem{theorem}{Theorem}

\newtheorem{lemma}{Lemma}

\newtheorem{proposition}{Proposition}

\newtheorem{probstat}{Problem Statement}

\begin{document}
\begin{frontmatter}

\title{Continuous Reachability Task Transition Using Control Barrier Functions\thanksref{footnoteinfo}}

\thanks[footnoteinfo]{This work was supported in part by the National Science Foundation under award \#1749357. C. Santoyo was supported by the NSF Graduate Research Fellowship Program under Grant No. DGE1650044.}

\author[FirstSecond]{Mohit Srinivasan}
\author[FirstSecond]{Cesar Santoyo} 
\author[Third]{Samuel Coogan}

\address[FirstSecond]{Georgia Institute of Technology, School of Electrical \& Computer Engineering, Atlanta, GA, 30318, USA, \\ Email: \{mohit.srinivasan, csantoyo\} @gatech.edu}
\address[Third]{Georgia Institute of Technology, School of Electrical \& Computer Engineering and School of Civil \& Environmental Engineering, Atlanta, GA, 30318, USA, \\ Email: sam.coogan@gatech.edu}

\begin{abstract}
In this paper, a method to achieve smooth transitions between sequential reachability tasks for a continuous time mobile robotic system is presented. Control barrier functions provide formal guarantees of forward invariance of safe sets and finite-time reachability and are able to enforce task execution. Barrier functions used in quadratic programs result in implementation of controllers with real-time performance guarantees. Existing approaches for multi-objective task execution using control barrier functions leverage discretely switched, sequential quadratic programs to achieve successive tasks. However, discrete switching can lead to control input discontinuities which can affect a robot's performance. Hence, we propose a method which ensures continuous transitions between sequential quadratic programs. In particular, a time varying component to the barrier function constraint is introduced which allows for a smooth transition between objectives. Robotic implementation results are also provided.
\end{abstract}

\begin{keyword}
Mobile robots, Autonomous robotic systems, Robotics technology
\end{keyword}

\end{frontmatter}

\section{Introduction}
\label{sec:intro}
Increased complexity of autonomous robotic systems and demands for safety guarantees have made it imperative that formal guarantees of safety and performance are a cornerstone of control synthesis. In particular, synthesized actions are often required to be continuous inputs which do not result in spasmodic motions. For example, in systems such as humanoid robots \cite{park_con_task_trans}, \cite{park_singularity}, it is critical that the control inputs to the manipulator joints are continuous in nature, and do not suffer from discontinuities. To that end, we present a framework which satisfies a system's performance and safety specifications while ensuring continuity in the control law when transitioning to another objective.

Control barrier functions (CBFs) used in conjunction with quadratic programs (QPs) lead to real-time implementation of controllers which satisfy safety and reachability requirements. For example, \cite{Li_hetero} utilize CBFs to ensure collision avoidance amongst teams of robots. \cite{ames2014cdcCBFs} employ CBFs to guarantee speed and car separation objectives in the context of adaptive cruise control. More recently, \cite{mohit_TRO} leveraged CBFs for motion planning specifications for teams of robots. \cite{pietro_darpa} execute a sequence of tasks and transition from an existing graph structure to the desired one using CBFs. In these papers, the authors transition discretely between tasks using sequential QPs. Each QP encodes a different barrier function constraint based on the task to be satisfied.

However, as is shown in \cite{park_con_task_trans}, sudden switching between constraints can lead to discontinuities in the control input which adversely affect the stability and functionality of the robot. Hence, in this paper, we address this issue by formulating a barrier function constraint with time dependent properties which allow for continuous transition between different tasks. More specifically, each barrier function is endowed with a time-varying transition coefficient which acts as a buffer signal that appropriately switches the constraint on or off. The authors in \cite{park_con_task_trans} use a similar time varying parameter to smoothly transition between different tasks in the context of humanoid robots and manipulators. However, the authors do not provide any formal guarantees on the satisfaction of the newly added task or the continuity of the proposed controller.

The contributions of this paper are threefold. First, we formulate a new barrier function constraint which allows for smooth transitions between different reachability tasks to be satisfied by a mobile robotic system while simultaneously satisfying safety constraints. With a simple example, we demonstrate that discrete switching between successive QPs leads to a discontinuity in the control law whereas the proposed constraint allows for a continuous transition of the control. Second, we provide an algorithm for online implementation of the proposed barrier function based controller. In addition, we demonstrate controller continuity. Lastly, we present experimental results which verify the proposed framework on a differential drive mobile robot.

The paper is organized as follows: Section~\ref{sec:math_bkg} provides mathematical background on control barrier functions, and the QP-based controller. Section~\ref{sec:mot_ex_prob_stat} details a motivating example which illustrates why a discrete transition between sequential QPs can lead to discontinuous control inputs, and also formalizes the problem statement addressed in this paper. Section~\ref{sec:main_results} discusses the proposed barrier function constraint which guarantees continuity of the control law. Section~\ref{sec:example} and \ref{sec:conclusion} describe the applicability of the proposed controller, and concluding remarks, respectively.

\section{Mathematical Background}
\label{sec:math_bkg}
Consider a control affine dynamical system of the form
\begin{align}
\label{eq:system}
    \dot x = f(x) + g(x)u \text{,}
\end{align}
where $x \in \mathcal{D} \subset \mathbb{R}^{n}$ is the state of the system, $u \in \mathbb{R}^{m}$ is the control input applied to the system, and $f$ and $g$ are locally Lipschitz continuous functions.

\subsection{Zeroing Control Barrier Functions (ZCBFs)}
Let $\mathcal{C} = \{ x \in \mathcal{D} \mid h(x) \geq 0 \}$ be a safe set defined as the super zero level-set of a continuously differentiable function $h : \mathcal{D} \rightarrow \mathbb{R}$. We define a continuous function $\mu : \mathbb{R} \rightarrow \mathbb{R}$ as a class $\kappa$ function if $\mu(0) = 0$ and it is strictly increasing.
\\
\begin{defin}[\cite{CBFs_tutorial}]
    \label{def:zcbf}
    A continuously differentiable function $h : \mathcal{D} \rightarrow \mathbb{R}$ is a zeroing control barrier function (ZCBF) if there exists a locally Lipschitz class $\kappa$ function $\mu$ such that for all $x \in \mathcal{D}$
    \begin{align*}
        \label{eq:zcbf}
        \sup\limits_{u \in \mathbb{R}^{m}} \{ L_{f}h(x) + L_{g}h(x) u + \mu(h(x)) \} \geq 0 \ ,
    \end{align*}
where $L_{f}h(x) = \frac{\partial h(x)}{\partial x}f(x)$, and $L_{g}h(x) = \frac{\partial h(x)}{\partial x}g(x)$.
\end{defin}

In particular, choosing control inputs from the set
\begin{equation*}
    \label{eq:inv_controls}
    \mathcal{U}_{\text{safe}}(x) = \{ u \in \mathbb{R}^{m} \mid L_{f}h(x) + L_{g}h(x) u + \mu(h(x)) \geq 0\}
\end{equation*}
renders $\mathcal{C}$ forward invariant. Since class $\mathcal{K}$ functions fall under the category of minimal functions, using the formalism in \cite{mcbfs}, we can guarantee forward invariance of $\mathcal{C}$.
\\
\begin{proposition}[Theorem 6, \cite{mcbfs}]
\label{prop:zcbf}
Let $\mathcal{C} \subset \mathcal{D}$ be a safe set defined as $\mathcal{C} = \{ x \in \mathcal{D} \,|\, h(x) \geq 0\}$ where $h : \mathcal{D} \rightarrow \mathbb{R}$, and suppose $x(0) \in \mathcal{C}$. If $h$ is a ZCBF, then any continuous feedback controller satisfying $u(x) \in \mathcal{U}_{\text{safe}}(x)$ for all $x \in \mathcal{C}$ renders the set $\mathcal{C}$ forward invariant.
\end{proposition}

\subsection{Finite-time Control Barrier Functions (FCBFs)}
Let $\Gamma = \{ x \in \mathcal{D} \mid h(x) \geq 0\}$ be a target set defined as the super zero level-set of a continuously differentiable function $h : \mathcal{D} \rightarrow \mathbb{R}$. \cite{li2018formally} introduce finite-time control barrrier functions (FCBFs) which guarantee finite-time convergence to a desired target set in the domain, as formalized next.
\\
\begin{defin}
\label{def:fcbf}
A continuously differentiable function $h : \mathcal{D} \rightarrow \mathbb{R}$ is a finite-time control barrier function (FCBF) if there exists parameters $\gamma >0 $ and $\rho \in [0,1 )$ such that for all $x \in \mathcal{D}$
\begin{equation*}
    \label{eq:fcbf}
     \sup\limits_{u \in \mathbb{R}^{m}}
    \{ L_{f}h(x) + L_{g}h(x) u + \gamma \cdot \sign(h(x)) \cdot |h(x)|^{\rho} \} \geq 0.
\end{equation*}
where $L_{f}h(x) = \frac{\partial h(x)}{\partial x}f(x)$, and $L_{g}h(x) = \frac{\partial h(x)}{\partial x}g(x)$.
\end{defin}

Given a target set $\Gamma$, choosing control inputs from the set
\begin{align*}
    \label{eq:fcbf_controls}
    \mathcal{U}_{\text{target}}(x) = \{ u \in \mathbb{R}^{m} &\mid L_{f}h(x) + L_{g}h(x) u + \nonumber \\
    &\gamma \cdot \sign(h(x)) \cdot |h(x)|^{\rho} \geq 0\}
\end{align*}
allows the system to converge to $\Gamma$ in finite time, as formalized in the following proposition.
\\
\begin{proposition}[Proposition III.1, \cite{li2018formally}]
\label{prop:fcbf}
Let $\Gamma \subset \mathcal{D}$ be a target set defined as $\Gamma = \{ x \in \mathcal{D} \mid h(x) \geq 0\}$ where $h : \mathcal{D} \rightarrow \mathbb{R}$. If $h$ is a FCBF, then, for any initial condition $x_{0} \in \mathcal{D} \backslash \Gamma$ and any continuous feedback control $u : \mathcal{D} \rightarrow \mathbb{R}^{m}$ satisfying $u(x) \in \mathcal{U}_{\text{target}}(x) $ for all $x \in \mathcal{D}$, the system will be driven to the set $\Gamma$ in a finite time $0 < T < \infty$; that is, $x(T)\in \Gamma$.
\end{proposition}

\subsection{Quadratic Program Based Controller}
Given a FCBF and/or a ZCBF, control synthesis can be encoded as a QP which is amenable to efficient online computation of feasible control inputs. In particular, for fixed $x \in \mathcal{D}$, the requirement that $u(x) \in \mathcal{U}_{\text{safe}}(x)$ and/or $u(x) \in \mathcal{U}_{\text{target}}(x)$ becomes a linear constraint and we define a minimum energy QP as
\begin{equation}
\begin{aligned}
\label{intro_qp}
& \underset{u \in \mathbb{R}^{m}}{\text{minimize}}
\quad ||u||_{2}^{2}\\
& \text{\quad s.t \quad \quad \quad} u \in \mathcal{U}_{\text{target}}(x) \text{ and/or }u \in \mathcal{U}_{\text{safe}}(x) \text{.}
\end{aligned}
\end{equation}

If feasible for all time, \eqref{intro_qp} returns the point-wise in time, minimum energy control action which guarantees that the system satisfies the safety constraints dictated by the ZCBF (Proposition~\ref{prop:zcbf}) and/or the reachability constraints dictated by the FCBF (Proposition~\ref{prop:fcbf}). While the QP can be constrained individually by either the reachability or safety constraints, we are specifically interested in achieving reachability tasks while simultaneously enforcing safety. 

\section{Motivating Example \& Problem Statement}
\label{sec:mot_ex_prob_stat}
Suppose the system of interest is an omnidirectional robot with single integrator dynamics $\dot{x} = u$, where $x \in \mathbb{R}^{2}$ is the state, and $u \in \mathbb{R}^{2}$ is the control input applied to the robot. Consider a domain $\mathcal{D} \subset \mathbb{R}^2$ containing two regions of interest- region $A$ and region $B$. Let $\mathcal{A} = \{x \in \mathcal{D} \mid h_{A}(x) \geq 0 \}$ and $\mathcal{B} = \{x \in \mathcal{D} \mid h_{B}(x) \geq 0 \}$, where $h_{A} : \mathbb{R}^2 \rightarrow \mathbb{R}$ and $h_{B} : \mathbb{R}^2 \rightarrow \mathbb{R}$ are continuously differentiable functions. Suppose the task to be satisfied by the robot is to visit region $A$ first followed by region $B$. Adhering to the methodology suggested in \cite{mohit_TRO}, one can formulate this task as a sequence of two QPs.

\textit{Quadratic Program 1: }The first QP is defined to be
\begin{equation*}
\begin{aligned}
& \underset{u \in \mathbb{R}^{2}}{\text{min}}
\quad ||u||_{2}^{2}\\
& \text{s.t \quad} \frac{\partial h_{A}(x)}{\partial x} \cdot u \geq -\gamma \cdot \sign(h_{A}(x)) \cdot |h_{A}(x)|^{\rho}
\end{aligned}
\end{equation*}
where $\gamma > 0$, and $\rho \in [0,1)$.
This QP is solved until the system reaches $\mathcal{A}$.

\textit{Quadratic Program 2: } Next, the QP constraint is switched to a new one to reflect the new task, i.e., convergence to region B. Hence, the second QP is defined to be
\begin{equation*}
\begin{aligned}
& \underset{u \in \mathbb{R}^{2}}{\text{min}}
\quad ||u||_{2}^{2}\\
& \text{s.t \quad} \frac{\partial h_{B}(x)}{\partial x} \cdot u \geq -\gamma \cdot \sign(h_{B}(x)) \cdot |h_{B}(x)|^{\rho}
\end{aligned}
\end{equation*}
where $\gamma > 0$, and $\rho \in [0,1)$.
\begin{figure}
    \centering
    \includegraphics[scale=.4]{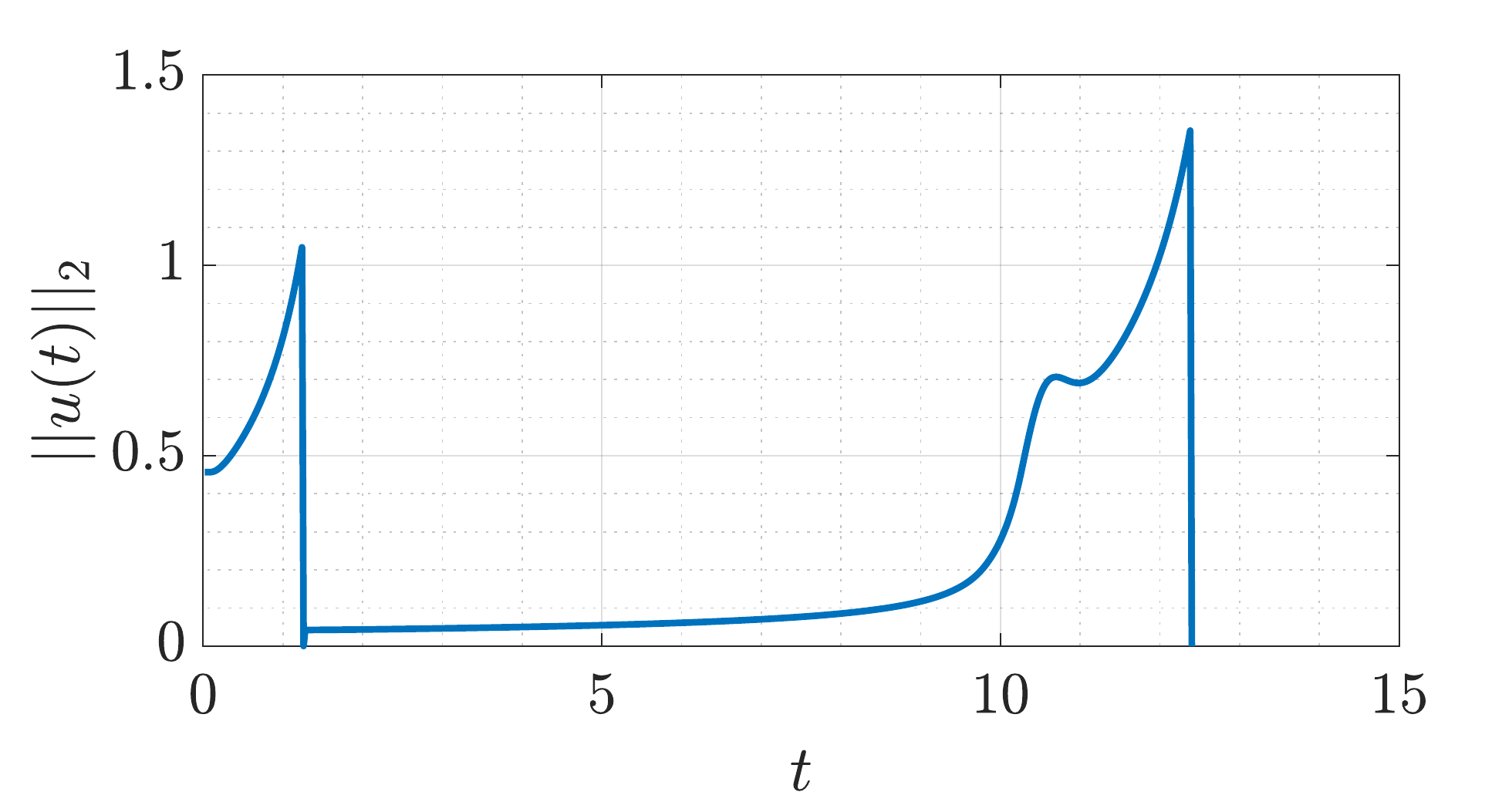}
    \caption{
    The control input generated for the sequence of tasks discussed in Example 1. The sudden switching of the reachability constraint from QP 1 to QP 2 yields a discontinuity in the control law as is seen here. In Section~\ref{sec:example}, we show that using our proposed framework, a continuous control input is obtained.}
    \label{fig: ctrl_discon}
\end{figure}

The sudden change in the constraint yields a discontinuous control law as illustrated in Fig~\ref{fig: ctrl_discon}. A similar example of this problem is discussed in \cite{park_con_task_trans} and  \cite{park_hqp}. Hence, discretely switching from one QP to the next creates discontinuities in the control input to the robot, which is undesirable. To address this problem, we propose a new barrier function constraint with time dependent coefficients which guarantee continuity of the control law. Instead of solving a sequence of QPs with discrete transitions, we propose a single QP which smoothly transitions between different barrier function constraints.

To that end, consider a continuous time mobile robotic system in control affine form as in \eqref{eq:system}. We assume that there exists a sequence of tasks to be executed by the system. Below, we provide a formal definition for a task.
\\
\begin{defin}
Given a target set $\Gamma \subset \mathcal{D}$ and a safety set $\Sigma \subset \mathcal{D}$, the task $\mathcal{T}_{(\Gamma, \Sigma)}$ is defined as the reachability problem requiring that the system reach $\Gamma$ in finite time, while staying in $\Sigma$ for all time.
\end{defin}

Define $\mathcal{T} = \{ \mathcal{T}_{(\Gamma_1, \Sigma)}, \dots, \mathcal{T}_{(\Gamma_n, \Sigma)} \}$ where each $\mathcal{T}_{(\Gamma_i, \Sigma)}$ is a task that the system must satisfy in finite time and each sequential task is distinct, i.e., $\mathcal{T}_{(\Gamma_i, \Sigma)} \neq \mathcal{T}_{(\Gamma_{i+1}, \Sigma)}$ for all $i$.

Informally, each task $\mathcal{T}_{(\Gamma_i, \Sigma)}$ consists of reachability and invariance constraints. Invariance constraints are assumed to be global constraints which do not change between consecutive tasks; however, the reachability constraints are distinct between tasks. These constraints are encoded in a QP which is solved until the system reaches the target set $\Gamma_i$. Suppose there exists $m$ finite-time barrier functions and each target set $\Gamma_i$ for the task $\mathcal{T}_{(\Gamma_i, \Sigma)}$ is characterized as the intersection of some subset of the $m$ barrier functions. Let $\mathcal{I} = \{ 1,2,\dots,m\}$ be the index set for the finite-time barrier functions. Given a target set $\Gamma_i$ for a task $i$, the following definition formalizes the finite-time barrier functions which characterize $\Gamma_i$.
\\
\begin{defin}[Reachability Set Map]
\label{def:reach_mapping}
The reachability set map $\Delta: \{ \Gamma_1, \Gamma_2, \ldots, \Gamma_n \} \rightarrow 2^{\mathcal{I}}$ yields the finite-time barrier functions characterizing the target set such that $\Gamma_i = \bigcap\limits_{\forall j \in \Delta(\Gamma_i)}  \{ x \in \mathcal{D} \mid h_{j}(x) \geq 0\}$ for all $i \in \{1,2,\dots, n\}$.
\end{defin}

Next, we formalize the problem statement addressed in this paper.
\\
\begin{probstat}
Given a system as in \eqref{eq:system}, and a sequence of tasks $\mathcal{T} = \{ \mathcal{T}_{(\Gamma_1, \Sigma)}, \dots, \mathcal{T}_{(\Gamma_n, \Sigma)}\}$, synthesize a continuous controller such that the system satisfies each $\mathcal{T}_{(\Gamma_i, \Sigma)}$ within a finite time, while smoothly transitioning between sequential tasks.
\end{probstat}

\section{Barrier Function Based Smooth Task Transition}
\label{sec:main_results}
Smoothly transitioning between tasks requires a transition function which gradually winds down the weight(s) of the constraint(s) corresponding to an accomplished task in the quadratic program (QP) while ramping up the weight(s) of the constraint(s) of the successive task. In this section, we propose a framework to achieve this. 

\subsection{QP Based Algorithm}
Consider a sequence of $n$ reachability tasks represented as $\mathcal{T} = \{ \mathcal{T}_{(\Gamma_{1}, \Sigma)},$ $\dots, \mathcal{T}_{(\Gamma_{n}, \Sigma)}\}$, and $m$ finite time barrier functions indexed by the set $\mathcal{I}$. Introduce transition function $\alpha_{i} : \mathbb{R}_{\geq 0} \rightarrow \mathbb{R}_{\geq 0}$ for all $i \in \mathcal{I}$. Then, for all $t \geq 0$, we formulate a time varying feedback control QP of the form
\begin{equation}
\begin{aligned}
\label{eq:QP_controller}
u^{*}(t, x) = & \underset{u \in \mathbb{R}^{m}}{\text{min}}
\quad ||u||_{2}^{2}\\
\text{s.t \quad}  &\sum\limits_{i = 1}^{m} \bigg\{ \alpha_{i}(t)(L_{f}h_{i}(x(t)) + L_{g}h_{i}(x(t))u(x(t))) \bigg\} + \\
& \sum\limits_{i=1}^{m} h_{i}(x(t)) \frac{\partial \alpha_{i}(t)}{\partial t} \\
& \geq - \gamma \cdot \tanh\bigg( -\ln\bigg( \sum\limits_{i=1}^{m} \exp(-\alpha_{i}(t) h_{i}(x(t))\bigg)\bigg)\\
& u \in \mathcal{U}_{\text{safe}}(x) \\
& ||u||_{\infty} \leq M, 
\end{aligned}
\end{equation}
where $\alpha_{i}(t)$ is determined from Algorithm~\ref{algo: alpha_function} for all $i \in \mathcal{I}$, and $\gamma > 0$. The first constraint captures the reachability part of each task specification, whereas the second constraint represents the safety requirements to be satisfied by the system. The last constraint represents actuator limits where $M > 0$. This optimization problem must be solved point-wise in time i.e. in a sampled-data fashion.

We construct Algorithm~\ref{algo: smooth_procedure} and Algorithm~\ref{algo: alpha_function}, which use \eqref{eq:QP_controller} to generate the control inputs required to satisfy the given sequence of reachability tasks for the system. Algorithm \ref{algo: smooth_procedure} utilizes the function characterized in Algorithm \ref{algo: alpha_function}. In particular, Algorithm \ref{algo: smooth_procedure} is executed for all $t \geq 0$. The transition functions $\alpha_i(t)$ for all $i \in \mathcal{I}$ and for all $t \geq 0$ are chosen as per Algorithm~\ref{algo: alpha_function} point-wise in time. The functions $\kappa^{\uparrow} : \mathbb{R}_{\geq 0} \rightarrow \mathbb{R}$ and $\kappa^{\downarrow} : \mathbb{R}_{\geq 0} \rightarrow \mathbb{R}$ are strictly increasing and decreasing continuously differentiable functions, respectively. Assuming $T_i$ is the time instant at which the robot reaches target set $\Gamma_{i}$, the functions $\kappa^{\uparrow}$ and $\kappa^{\downarrow}$ are chosen such that $\kappa^{\uparrow}(t - T_i)\mid_{t = T_i} = 0$ and $\kappa^{\downarrow}(t - T_i)\mid_{t = T_i} = 1$. Choices for $\kappa^{\uparrow}$ and $\kappa^{\downarrow}$ include functions such as sine, cosine, hyperbolic tangent, sigmoid, etc.
\\
\begin{algorithm}
\begin{algorithmic}[1]
    \Procedure{}{}
    \hspace*{\algorithmicindent} \\
    \textbf{Input : $h_{i}$ for all $i \in \mathcal{I}$, $\gamma > 0$}
    \For{$i \in \{ 1,\ldots,n\}$} \Comment{Loop through each task}
        \State $T_i \gets 0$
        \State $s \gets 0$ \Comment{Flag variable to indicate transition}
        \While{$x \not\in \Gamma_i$} \Comment{Reachability Phase}
            \State \textproc{Compute-$\alpha(t, T_i, x,i,s)$}
            \State Solve the QP \eqref{eq:QP_controller}
            \State Apply $u(t, x)$ to the system
        \EndWhile
        \State $T_{i} \gets t$ \Comment{Record time instant when $x(T_i) \in \Gamma_i$}
        \State $s \gets 1$
        \While{$\alpha_k < 1$ $ \And$ $\alpha_j > 0 \ \ \forall$ $k \in \Delta(\Gamma_{i+1})$, $j \in \Delta(\Gamma_i)$} \Comment{Transition Phase}
            \State \textproc{Compute-$\alpha(t,T_i, x,i,s)$}
            \State Solve the QP \eqref{eq:QP_controller}
            \State Apply $u(t,x)$ to the system
        \EndWhile
    \EndFor
    \EndProcedure
    \caption{Smooth Transition Between Sequential Reachability Tasks}
    \label{algo: smooth_procedure}
\end{algorithmic}
\end{algorithm}
\begin{algorithm}
    \begin{algorithmic}[1]
        \Function{Compute-$\alpha$}{$t$, $T_i$, $x$, $i$, $s$}
        \If{$s = 0$}
            \State $\alpha_j \gets 1$ for all $j \in \Delta(\Gamma_i)$
            \State $\alpha_k \gets 0$ for all $k \in \mathcal{I}\setminus\Delta(\Gamma_i)$
        \ElsIf{$s = 1$}
            \State $\alpha_j \gets \kappa^{\downarrow}(t-T_i)$ for all $j \in \Delta(\Gamma_i)$
            \State $\alpha_l \gets \kappa^{\uparrow}(t-T_i)$ for all $l \in \Delta(\Gamma_{i+1})$
        \EndIf 
        \EndFunction
        \caption{Compute Transition Function}
        \label{algo: alpha_function}
    \end{algorithmic}
\end{algorithm}

\subsection{Continuity of QP-based Controller \& Reachability Task Satisfaction Guarantee}
The following theorem provides a composite barrier function constraint which allows for smooth task transition.
\\
\begin{lemma}
\label{lemma:sequence_barrier}
Consider a task $\mathcal{T}_{(\Gamma, \Sigma)}$ with the target set $\Gamma$ defined as $\Gamma = \bigcap\limits_{i \in \mathcal{P}}\{ x \in \mathcal{D} \mid h_{i}(x) \geq 0\}$ where $\mathcal{P} = \{ 1,2,\ldots,k\}$, with each barrier function $h_{i}$ bounded. That is, $h_{i}(x) \leq M_i$ for all $x \in \mathcal{D}$ where $M_i > 0$. If there exists a continuous controller $u : \mathcal{D} \rightarrow \mathbb{R}^{m}$ such that for all $x \in \mathcal{D}$ and for all $t \geq 0$,
\begin{multline}
\label{eq:smooth_cbf}
\sum\limits_{i \in \mathcal{P}} \bigg\{ (L_{f}h_{i}(x(t)) + L_{g}h_{i}(x(t))u(x(t))) \bigg\} \\
\geq - \gamma \cdot \tanh\bigg( -\ln\bigg( \sum\limits_{i \in \mathcal{P}} \exp(-h_{i}(x(t))\bigg)\bigg)
\end{multline}
then there exists a time instance $0 < T < \infty$ such that $x(T) \in \Gamma$.
\end{lemma}
\begin{pf}\let\qed\relax
By contradiction, suppose for some $x(0) \in \mathcal{D} \backslash \Gamma$, the control law that satisfies \eqref{eq:smooth_cbf} is such that there does not exist a time $0 < T < \infty$ so that $x(T) \in \Gamma$.

Hence we have $\min\limits_{i \in \mathcal{P}}( \{ h_{i}(x(t))\}) < 0$. 
From \cite{boyd}, p. 72, we therefore have
\begin{equation*}
    -\ln\bigg( \sum\limits_{i \in \mathcal{P}} \exp(-h_{i}(x(t)))\bigg) < 0
\end{equation*}
for all $t \geq 0$. Since $\tanh(x) < 0$ for all $x < 0$, we have for all $t \geq 0$
\begin{equation*}
    \tanh\bigg( -\ln\bigg( \sum\limits_{i \in \mathcal{P}} \exp(-h_{i}(x(t))\bigg)\bigg) = -\beta < 0
\end{equation*}

Observe that the inequality \eqref{eq:smooth_cbf} can be rewritten as
\begin{align}
\label{eq:time_derivative}
    \frac{d}{dt}\sum\limits_{i \in \mathcal{P}} h_{i}(x(t)) \geq \gamma \cdot \beta
\end{align}
By integration of \eqref{eq:time_derivative} using the fundamental theorem of calculus and from \eqref{eq:smooth_cbf}, we thus get
\begin{align*}
    \sum\limits_{i \in \mathcal{P}} h_{i}(x(t)) \geq \gamma \cdot \beta \cdot t +
    \sum\limits_{i \in \mathcal{P}} h_{i}(x(0))
\end{align*}

Observe that as $t \rightarrow \infty$, $\sum\limits_{i \in \mathcal{P}} h_{i}(x(t)) \rightarrow \infty$. However, this is a contradiction since the barrier functions are bounded, and hence we have $\sum\limits_{i \in \mathcal{P}} h_{i}(x(t)) < \sum\limits_{i \in \mathcal{P}} M_i$ for all $t \geq 0$. Hence there exists a $0 < T < \infty$ such that $x(T) \in \Gamma$.
\QEDA
\end{pf}
The following theorem reformulates the constraint \eqref{eq:smooth_cbf} taking into account the transition periods of the transition functions given a sequence of $n$ tasks to be executed by the system.
\\
\begin{theorem}
\label{thm:ntask_barrier}
Consider $m$ bounded finite-time barrier functions, i.e., $h_{i}(x) \leq M_i$ for all $x \in \mathcal{D}$ and where $M_i > 0$ for all $i \in \{1,2,\dots,m \}$. Given a sequence of tasks $\mathcal{T} := \{ \mathcal{T}_{(\Gamma_1, \Sigma)}, \ldots, \mathcal{T}_{(\Gamma_n, \Sigma)}\}$ with the corresponding transition functions $\alpha_{j}(t)$ chosen according to Algorithm \ref{algo: alpha_function} for all $j \in \Delta(\Gamma_i)$, for all $i \in \{ 1,\ldots,n\}$, and for all $t \geq 0$, if $u : \mathbb{R}_{\geq 0} \times \mathcal{D} \rightarrow \mathbb{R}^{m}$ is a continuous controller such that for all $x \in \mathcal{D}$ and $t \geq 0$, we have
\begin{multline}
\label{eq:smooth_cbf_2}
\sum\limits_{i = 1}^{m} \bigg\{ \alpha_{i}(t)(L_{f}h_{i}(x(t)) + L_{g}h_{i}(x(t))u(t, x(t))) \bigg\} + \\
\sum\limits_{i=1}^{m} h_{i}(x(t)) \frac{\partial \alpha_{i}(t)}{\partial t} \\
\geq - \gamma \cdot \tanh\bigg( -\ln\bigg( \sum\limits_{i=1}^{m} \exp(-\alpha_{i}(t) h_{i}(x(t))\bigg)\bigg),
\end{multline}
then there exists a sequence of finite time instances $0 < T_1 < T_2 < \ldots < T_n < \infty$ such that $x(T_i) \in \Gamma_i$ for all $i \in \{ 1,\ldots,n\}$ i.e. the task sequence $\mathcal{T}$ is solved.
\end{theorem}
\begin{pf} \let\qed\relax
The proof is analogous to the proof of Lemma~\ref{lemma:sequence_barrier}. Consider the first task $\mathcal{T}_{(\Gamma_1, \Sigma_1)}$. From Lemma~\ref{lemma:sequence_barrier} we know that there exists a finite time $0 < T_1 < \infty$ such that $x(T_1) \in \Gamma_1$. Suppose by contradiction, for some $x(T_1) \in \mathcal{D} \backslash \Gamma_2$, the constraint \eqref{eq:smooth_cbf_2} is satisfied for all $t \geq T_1$ but there does not exist a finite time $0 < T_2 < \infty$ such that $x(T_2) \in \Gamma_2$. Hence we have that $\min\limits_{i \in \{1,2,\ldots, m\}}( \{ \alpha_{i}(t) h_{i}(x(t))\}) = \min\limits_{i \in \Delta(\Gamma_2)}( \{ \alpha_{i}(t) h_{i}(x(t))\}) < 0$.
Following a proof methodology similar to Lemma~\ref{lemma:sequence_barrier}, we can prove that there exists a finite time $0 < T_2 < \infty$ such that $x(T_2) \in \Gamma_2$. Following a successive proof by induction for tasks $3$ to $n$, we can thus conclude that there exists a sequence of finite time instances $0 < T_1 < T_2 < \ldots < T_n < \infty$ such that $x(T_i) \in \Gamma_i$ for all $i \in \{1,2,\ldots,n \}$ i.e. the task sequence $\mathcal{T}$ is solved. \QEDA
\end{pf}

Since we are interested in guaranteeing continuity of the controller, the following theorem proves that the proposed controller \eqref{eq:QP_controller} as used in Algorithm~\ref{algo: smooth_procedure} is continuous.
\\
\begin{theorem}
If Algorithm 1 is feasible for all $t \geq 0$, then the control input, computed by solving \eqref{eq:QP_controller}, applied to the system is continuous.
\end{theorem}
\begin{pf} \let\qed\relax
Consider an indicator variable for the time, $\theta$ with $\dot{\theta} = 1$. Considering the new state of the system as $\hat{x} = \begin{bmatrix} x & \theta \end{bmatrix}^{T}$, the constraint \eqref{eq:smooth_cbf_2} can be reformulated as
\begin{multline*}
\sum\limits_{i = 1}^{m} \bigg\{ \alpha_{i}(\hat{x})(L_{f}h_{i}(\hat{x}) + L_{g}h_{i}(\hat{x})u(\hat{x})) \bigg\} + \sum\limits_{i = 1}^{m} h_{i}(\hat{x}) \frac{\partial \alpha_{i}(\hat{x})}{\partial \theta} + \\
\gamma \cdot \tanh \bigg( -\ln\bigg( \sum\limits_{i=1}^{m} \exp(-\alpha_{i}(\hat{x}) h_{i}(\hat{x}))\bigg) \bigg) \geq 0.
\end{multline*}
The reformulated inequality is now the barrier function constraint \eqref{eq:smooth_cbf} for the new time invariant system with the augmented state $\hat{x}$. Similarly, the constraint $u(x) \in \mathcal{U}_{\text{safe}}(x)$ can be reformulated in terms of the new state. These constraints are quasi-convex in the control $u$ for all $\hat{x} \in \mathbb{R}^{n+1}$ and the cost is quasi-convex. In addition, the input is constrained over a compact set (third constraint in \eqref{eq:QP_controller}). However, since $\alpha_{i}(\hat{x})$ for all $i \in \mathcal{I}$ is chosen point-wise in time as per Algorithm~\ref{algo: alpha_function}, the continuity of the controller must be established by analyzing the transition phase, reachability phase, and the time instant at which the transition between phases occurs as per Algorithm~\ref{algo: smooth_procedure}. Since all the assumptions of Proposition 8 in \cite{mcbfs} are satisfied, the controller computed from Algorithm~\ref{algo: smooth_procedure} is continuous for the entirety of the transition and the reachability phase. Since the functions $\kappa{\uparrow}$ and $\kappa^{\downarrow}$ are chosen such that, $\kappa^{\uparrow}(t - T_i)\mid_{t = T_i} = 0$ and $\kappa^{\downarrow}(t - T_i)\mid_{t = T_i} = 1$, the constraint \eqref{eq:smooth_cbf_2} does not suffer from discontinuities when switching between the reachability and the transition phase. Thus, the controller computed as per Algorithm~\ref{algo: smooth_procedure} is continuous for all $t \geq 0$. \QEDA
\end{pf}

\section{Robotic Implementation}
\label{sec:example}
In this section, we present the implementation\footnote{\url{https://github.com/gtfactslab/2020_ContinuousTaskBarriers}} of the theoretical framework discussed in Section \ref{sec:main_results} in the Georgia Tech Robotarium by \cite{pickem}.

Consider a differential drive mobile robot with dynamics
\begin{align*}
    \dot{x} &= v \cdot cos(\phi) \\
    \dot{y} &= v \cdot sin(\phi) \\
    \dot{\phi} &= \omega
\end{align*}
where $x \in \mathbb{R}$ and $y \in \mathbb{R}$ are the position coordinates of the robot, $\phi \in [-\pi, \pi)$ is the orientation, $v \in \mathbb{R}$ is the linear velocity input, and $\omega$ is the angular velocity input. Denote $\tilde{x} = \begin{bmatrix} x & y & \phi \end{bmatrix}^{T}$, $\hat{x} = \begin{bmatrix} x & y \end{bmatrix}^{T}$, and $\mathcal{D} \subset \mathbb{R}^{3}$. Then we have $f(\tilde{x}) = 0$ and $g(\tilde{x}) = \begin{bmatrix} \cos(\phi) & 0 \\ \sin(\phi) & 0 \\ 0 & 1\end{bmatrix}$. Let $u = \begin{bmatrix} v & \omega \end{bmatrix}^{T}$ be the input vector to the robot. The experiment was conducted on the Robotarium multi-robot testbed at Georgia Tech. For more details on the hardware specifications and platform setup, please refer \cite{pickem}. Since the differential drive robot model is non-holonomic which leads to controllability issues, we use a technique known as the Near Identity Diffeomorphism (NID) to control the robot, as detailed in \cite{pickem}.

\begin{figure}
    \centering
    \includegraphics[height=5cm,width=\columnwidth]{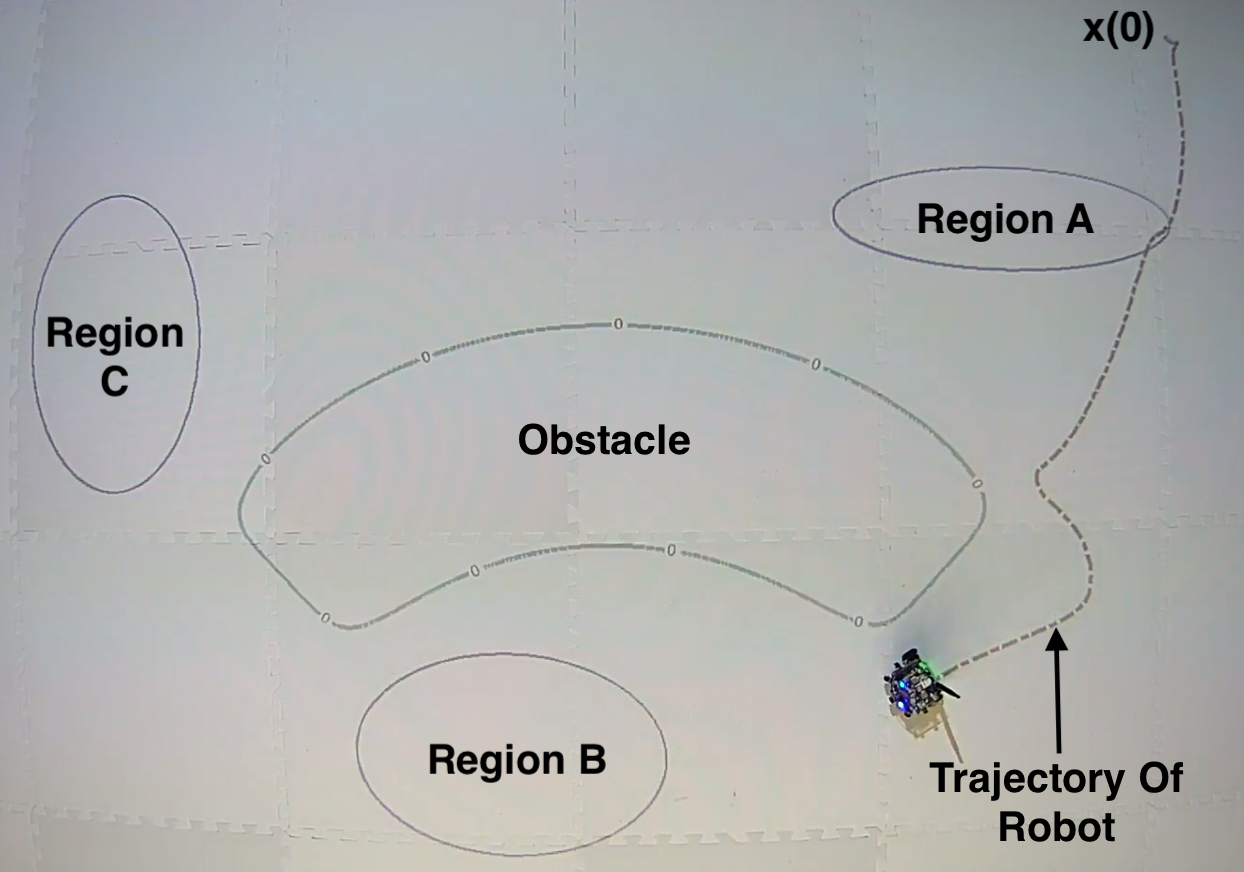}
    \caption{A still shot of the implementation of Algorithm~~\ref{algo: smooth_procedure} conducted on the Robotarium. The robot first visits region A, followed by region B, and lastly region C, while avoiding the obstacle.}
    \label{fig:domain}
\end{figure}

The workspace consists of four regions of interest - three goal regions (A, B and C) and an obstacle O. Fig.~\ref{fig:domain} illustrates the domain of interest. Here, the robot must initially visit region A, followed by region B and ultimately region C while avoiding the obstacle. The sequence of tasks is given by $\mathcal{T} = \{ \mathcal{T}_{(\Gamma_A, \Sigma_O)}, \mathcal{T}_{(\Gamma_B, \Sigma_O)}, \mathcal{T}_{(\Gamma_C, \Sigma_O)}\}$, where $\Gamma_A = \{ \tilde{x} \in \mathcal{D} \mid h_{1}(\tilde{x}) \geq 0\}$, $\Gamma_B = \{\tilde{x} \in \mathcal{D} \mid h_{2}(\tilde{x}) \geq 0\}$, $\Gamma_C = \{ \tilde{x} \in \mathcal{D} \mid h_{3}(\tilde{x}) \geq 0\}$, and $\Sigma_O = \{ \tilde{x} \in \mathcal{D} \mid h_{4}(\tilde{x}) \geq 0\}$, where each $h_{i} : \mathcal{D} \rightarrow \mathbb{R}$ for all $i \in \{1,2,3,4 \}$ is a continuously differentiable function. For regions A, B and C, the functions are given by $h_{1}(\tilde{x}) = 1 - (\hat{x} - C_A)^{T} P_A (\hat{x} - C_A)$, $h_{2}(\tilde{x}) = 1 - (\hat{x} - C_B)^{T} P_B (\hat{x} - C_B)$ and $h_{3}(\tilde{x}) = 1 - (\hat{x} - C_C)^{T} P_C (\hat{x} - C_C)$ where $P_A$, $P_B$, and $P_C$ are diagonal matrices with the inverses of the dimensions of the ellipsoids as the non-zero entries, and $C_A$, $C_B$ and $C_C$ are the centers of each ellipsoidal region. The obstacle is modeled using a weighted polar $L_p$ function as described in \cite{mohit_CDC2019} with parameters $p = 6$, $\sigma = (0.7, 0.2)$, $\theta_{\kappa} = \frac{\pi}{2}$, $c = 1$ and $\theta_0 = \sign(\kappa)\cdot \frac{\pi}{2}$.

At each time step $t$, Algorithm 1 is executed and the control input is applied to the robot. The QP that is solved to execute the entire sequence of tasks is 
\begin{equation*}
\begin{aligned}
u^{*}(t, \tilde{x}) = & \underset{u \in \mathbb{R}^{2}}{\text{min}}
\quad ||u||_{2}^{2}\\
& \text{s.t } \sum\limits_{i = 1}^{3} \bigg\{ \frac{d \alpha_{i}(t) h_{i}(\tilde{x}(t))}{dt} \bigg\}\\
& \geq -\gamma \cdot \tanh\bigg( -\ln\bigg( \sum\limits_{i=1}^{3} \exp(-\alpha_{i}(t) h_{i}(\tilde{x}(t))\bigg)\bigg) \\
& L_{f}h_{4}(\tilde{x}(t)) + L_{g}h_{4}(\tilde{x}(t)) u(t, \tilde{x}(t)) \geq - \gamma h_{4}(\tilde{x}(t))^{3} \\
& ||u||_{\infty} \leq 10
\end{aligned}
\end{equation*}
\begin{figure}
    \centering
    \includegraphics[height=5cm,width=\columnwidth]{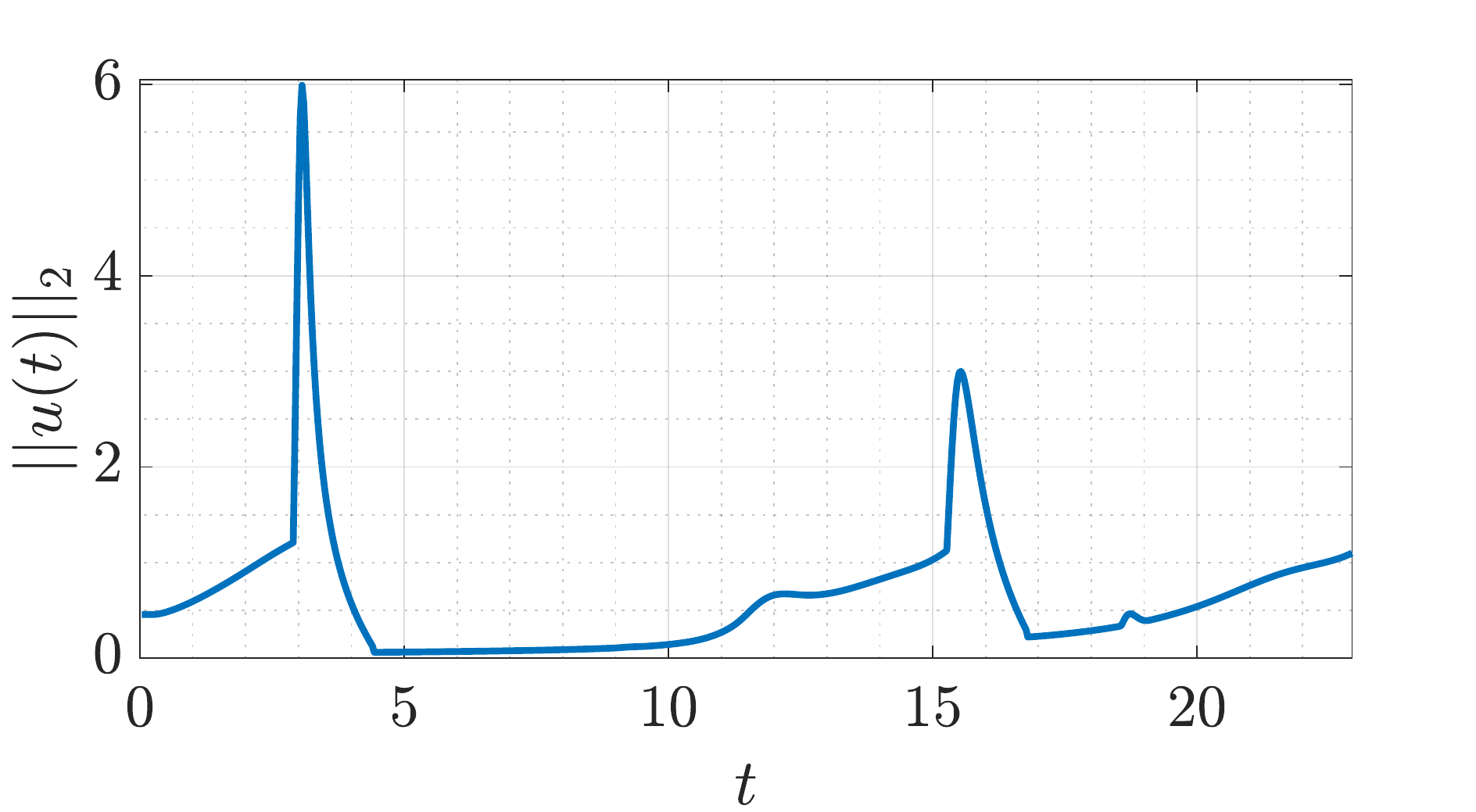}
    \caption{
    The control input $u$ generated in the Robotarium simulator for the task specification as discussed in the experimental setup. Observe that the control generated by our framework is continuous, which is in contrast to Fig~\ref{fig: ctrl_discon}.}
    \label{fig:ssol_space}
\end{figure}
where $L_{f}h_{i}(\tilde{x}(t)) = \frac{\partial h_{i}(\tilde{x}(t))}{\partial \tilde{x}} f(\tilde{x}) = 0$ and $L_{g}h_{i}(\tilde{x}(t)) = \frac{\partial h_{i}(\tilde{x}(t))}{\partial \tilde{x}} g(\tilde{x})$ for all $i \in \{ 1,2,3,4\}$.
The average time to solve the QP was between 3ms to 7ms. The increasing and decreasing functions in Algorithm~\ref{algo: alpha_function} are chosen as $\kappa^{\uparrow} = \sin^{2}(t-T_i)$ and $\kappa^{\downarrow}(t) = \cos^{2}(t-T_i)$ respectively, where $T_i$ is the recorded time instant at which the robot reaches region $i$. Fig.~\ref{fig:ssol_space} demonstrates the continuous change in the solution space as the robot is executing the QP. Observe that the discontinuous switching in the control input is avoided due to the modified constraint in \eqref{eq:smooth_cbf_2}. This is in contrast to the discontinuities shown in Fig~\ref{fig: ctrl_discon} where the traditional discrete QP switching method was used. A video of the implementation is provided.\footnote{Video of the implementation- \url{https://youtu.be/eKhXiJkQH8w}}

\section{Conclusion}
\label{sec:conclusion}
We proposed a control barrier function based method to ensure smooth transition between different reachability tasks for a control-affine robotic system. A new composite barrier function constraint was introduced by endowing individual barrier functions with time varying transition functions, which allow for smooth transitions between different objectives. In order to facilitate real-time implementation capabilities, we proposed an algorithm which incorporates the proposed barrier function constraint. Lastly, we proved that the proposed controller is continuous. Robotic implementation results are also provided which validate the proposed theory.

\bibliography{ifacconf}
\end{document}